# Rare-earth-free ferrimagnetic Mn$_4$N sub-20 nm thin films as high-temperature spintronic material

W. Zhou[1,a)], C.T. Ma[1], T.Q. Hartnett[2], P.V. Balachandran[2], S.J. Poon[1,a)]

[1] *Physics Department, University of Virginia, Charlottesville, Virginia,22903, USA*

[2] *Material Science Engineering, University of Virginia, Charlottesville, Virginia,22903, USA*

Ferrimagnetic alloy thin films that exhibit perpendicular (out-of-plane) magnetic anisotropy (PMA) with low saturation magnetization, such as GdCo and Mn$_4$N, were predicted to be favorable for hosting small Néel skyrmions for room temperature applications. Due to the exponential decay of interfacial Dzyaloshinskii-Moriya interaction (DMI) and the limited range of spin-orbit-torques, which can be used to drive skyrmion motion, the thickness of the ferrimagnetic layer has to be small, preferably under 20 nm. While there are examples of sub-20 nm, rare earth-transition metal (RE-TM), ferrimagnetic thin films fabricated by sputter deposition, to date rare-earth-free sub-20 nm Mn$_4$N films with PMA have only been reported to be achieved by molecular beam epitaxy, which is not suitable for massive production. Here we report the successful thermal growth of sub-20 nm Mn$_4$N films with PMA at 400-450 °C substrate temperatures on MgO substrates by reactive sputtering. The Mn$_4$N films were achieved by reducing the surface roughness of MgO substrate through a high-temperature vacuum annealing process. The optimal films showed low saturation magnetization ($M_s$ = 43 emu/cc), low magnetic anisotropy energy (0.7 Merg/cc), and a remanent magnetization to saturation magnetization ratio ($M_r/M_s$) near 1 at room temperature. Preliminary ab-initio density functional theory (DFT) calculations have confirmed the ferrimagnetic ground state of Mn$_4$N grown on MgO. The magnetic properties, along with the high thermal stability of Mn$_4$N thin films in comparison with RE-TM thin films, provide the platform for future studies of practical skyrmion-based spintronic materials.

[a)] Authors to whom correspondence should be addressed: wz8he@virginia.edu and sjp9x@virginia.edu

As industry rapidly transitions to using big data in their operations and decision making, there is an urgent need to develop technologies that accommodate the increasing requirements of high-density data storage [1]. One promising candidate that has received increasing attention is using magnetic skyrmions as information carriers and manipulating them with current by spin-orbit torque (SOT) for logic or memory operations [2,3] (e.g. racetrack memories). Magnetic skyrmions are swirling spin configurations of neighbor atoms in a magnetic material with topologic protection. The size of the skyrmion could be as small as a few nanometers [4] and manipulation of skyrmions could be energy efficient [5,6] in high-density data storage applications [7,8]. To date, most RT skyrmions have been discovered in ferromagnetic (FM) multilayer stacks [7-10]. The use of ferromagnets as skyrmions have certain shortcomings. For instance, ferromagnets suffer from large stray fields and large saturation magnetization ($M_s$), which makes the region of parameter space for small skyrmion less accessible, especially at room temperature (RT) [11].

Unlike ferromagnets, the ferrimagnetic counterparts have small stray fields and the low Ms, which makes them suitable for hosting small RT skyrmions [11]. This makes ferrimagnet a promising candidate material for high-density data storage applications. One prototypical example is the amorphous GdCo ferrimagnetic thin film. Small (10–30 nm) room temperature skyrmions have been reported in Pt/GdCo (6 nm)/TaO$_x$ [12]. While the amorphous GdCo ferrimagnet shows promising characteristics relative to the ferromagnets, it suffers from poor thermal stability. It has been shown that in amorphous GdCo, PMA is lost after annealing at 300-400 °C [13], the temperature range applied in complementary metal–oxide–semiconductor (CMOS) fabrication [14]. Since Néel skyrmions can only exist in a heterostructure with PMA, the poor thermal stability of RE-TM films will have a deleterious effect in both the fabrication and performance of amorphous RE-TM in devices that leverage skyrmions for storage technology. One of the potential solutions to this problem is to explore crystalline ferrimagnets with PMA synthesized by deposition at high-temperature near 400 °C, thus ensuring compatibility with conventional CMOS processing.

In addition to the intrinsic effects that we have discussed thus far, film thickness is an important extrinsic effect that impacts the overall device performance. The importance of film thickness applies to both ferro- and ferrimagnets. It is now well-established that the interfacial Dzyaloshinskii-Moriya interaction (DMI), which stabilizes the magnetic skyrmions in multilayers and heterostructures, decays exponentially with increasing film thickness [15-17]. Further, the SOT scales inversely with the thickness [18]. This necessitates the growth of thin film magnets with sub-20 nm thickness, for the realization of small skyrmions in practical applications [12,19,20].

Anti-perovskite $Mn_4N$ has been known as a crystalline, rare-earth-free ferrimagnetic material. In $Mn_4N$, the Mn-atoms have a face-centered cubic structure with one N-atom at the body center. A schematic of the unit cell is shown in Figure 1. The Mn-atoms at the corner and the face center have inequivalent magnetic moments and are ferrimagnetically coupled [21]. Although the easy axis of bulk $Mn_4N$ is along the [111] direction, PMA is repeatedly and reproducibly observed in crystalline $Mn_4N$ films [22-28]. As a result, ferrimagnetic $Mn_4N$ thin films have also attracted increasing interest in spintronics applications. Compared to the amorphous RE-TM ferrimagnetic alloys, the $Mn_4N$ system has better thermal stability for two main reasons. First, in most thin film studies, it has been shown that the anti-perovskite crystal structure is formed at 400-450 °C. No loss of PMA is reported in thin films after annealing, which is an encouraging outcome. Second, there is no known evidence for any structural phase transformation on cooling to room temperature. Thus, we interpret that the anti-perovskite crystal structure is tolerant of high-temperature device fabrication processes (unlike the GdCo amorphous alloys). As noted earlier, the emergence of PMA is an important magnetic property for its use in spintronic applications. One of the plausible reasons for the PMA in $Mn_4N$ thin films could be attributed to the deviation of the out-of-plane lattice constant, c, to the in-plane lattice constant, a, (c/a) ratio from 1, [22-28] due to the in-plane epitaxial strain. A recent study also showed that the anisotropy energy is correlated to the c/a ratio [27]. We note that further studies are warranted to understand the interplay of magnetic and strain effects on the PMA and this is beyond the scope of this paper.

Several groups have grown crystalline $Mn_4N$ epitaxial films (30-100 nm) on MgO, $SrTiO_3$, or $LaAlO_3$ substrates by magnetron sputtering [22,23], molecular beam epitaxy (MBE) [24-27], and pulsed laser deposition (PLD) [28], and they reported similar c/a ratios of ~0.99. The reported uniaxial magnetic anisotropy constant ($K_u$) and $M_s$ of those $Mn_4N$ films were about 0.5–1 Merg/cc and 50–100 emu/cc, respectively [22-28], comparable with the data observed for amorphous GdCo thin films ($K_u$ ~ 0.25 Merg/cc and $M_s$ ~ 50 emu/cc) [12].

To date, however, only a few groups have grown sub-20 nm $Mn_4N$ thin films with good magnetic hysteresis (M(H)) loops, which has remanent magnetization $M_r$ to $M_s$ ratio larger than 0.5 ($M_r/M_s > 0.5$), by MBE [26,27]. For those reported $Mn_4N$ film by sputtering, the thicknesses were 30 nm [22] to 100 nm [23], some of them even up to hundreds nanometers [29,30]. Similar results on sputter deposited sub-20 nm $Mn_4N$ thin film have not been reported. Compared to MBE, sputter deposition is a more widely adopted method in CMOS technology. In this work, we have epitaxially grown sub-20 nm $Mn_4N$ thin films on MgO (001) substrate with PMA by reactive sputtering. The effect of MgO substrate morphology on the quality of $Mn_4N$ film is studied. We then compare the magnetic properties of the $Mn_4N$ films on MgO substrate from experiment and first principles-based density functional theory (DFT) calculations. The main contribution of this paper lies in the demonstration of the growth of high magnetic quality sub-20 nm crystalline $Mn_4N$ films on MgO substrate using reactive sputtering that is more promising for scale-up production.

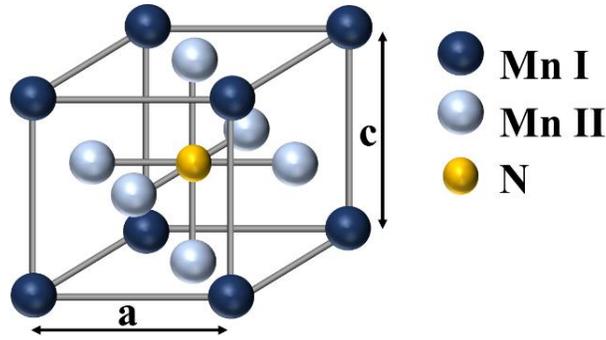

Figure 1. Mn$_4$N crystal structure

Mn$_4$N thin films with nominal thicknesses of ~15 nm and 10 nm were deposited on MgO(001) 5x5x0.5 mm substrates by reactive rf-sputtering at 400 °C and 450 °C substrate temperatures and a base pressure of 7x10$^{-8}$ Torr. The flow rates of Ar and N$_2$ gases were controlled by a mass flow meter and we maintained a flow rate Ar:N$_2$ ratio of 93:7. A 3 nm Pt capping layer was deposited on Mn$_4$N layer at room temperature to prevent oxidization. Before loading the MgO (001) substrate into the vacuum chamber, it was wet-cleaned with 2% diluted Hellmanex III alkaline detergent, acetone, and isopropanol, sealed in a vacuum tube with pressure 30 mTorr. The MgO substrate was baked at high temperatures 1,000 °C or 1,100 °C for 4 hours. Then, the substrates were annealed inside the chamber at 500 $^0$C for one hour to remove the surface contaminations. The Mn target was pre-sputtered with Ar gas for 20 minutes to remove the surface oxide. Surface roughness of the MgO substrate was measured using atomic force microscopy (AFM). The deposition rate was measured by X-ray reflectometry (XRR). The film compositions were determined with X-ray photoelectron spectroscopy (XPS). The film thickness was verified by XRR with an XRR simulation/calculation program (Rigaku, GXRR) [31].

The epitaxial growth of Mn$_4$N crystal layer was demonstrated using X-ray diffraction (XRD) with Cu-K$\alpha$ radiation. The magnetic properties of the samples were measured at room temperature with vibrating sample magnetometry (VSM). The diamagnetic component of the substrate was deduced from the slope of raw M-H curves at large H region and subtracted from the raw data. K$_u$ was calculated from effective anisotropy K with Eq. (1) and (2)

$$K_u = K + 2\pi M_s^2 \quad (1)$$

$$K = \int_0^\infty \left(M_{easy}(H) - M_{hard}(H)\right)dH \quad (2)$$

Where $M_{easy}$ and $M_{hard}$ are the magnetization in out-of-plane applied field and in-plane applied field, respectively.

*Ab-initio* electronic structure calculations were carried out in the Density Functional Theory (DFT) framework using the plane-wave pseudopotential Quantum ESPRESSO code [32, 33]. Core and valence electrons were treated using the ultrasoft pseudopotential method [34,35]. The exchange-correlation functionals were described using the Perdew-Burke-Ernzerhof parameterization of the generalized gradient approximation modified for solids (PBEsol) [36]. The plane-wave cutoff energy was set to 60 Ry and a Γ-centered Monkhorst-Pack *k*-point mesh of 12x12x12 was used to sample the Brillouin Zone [37]. Lattice parameters were fixed to experimentally measured value ($c/a = 0.987$) and $M_s$ was calculated using the formula, $M_s = \frac{|\mu_{total}|}{V}$, where $|\mu_{total}|$ is the absolute value of the total magnetization from DFT calculations given in Bohr magnetons and $V$ is the unit cell volume [38]. Self-consistent spin-polarized calculations were performed for collinear spin structures with no spin-orbit coupling term in the Hamiltonian. The Mn-atoms (Mn I) located in the cell corners with coordinates (0, 0, 0) were ferromagnetically coupled to Mn-atoms (Mn IIa) located at (0.5, 0.5, 0). Both Mn I- and Mn IIa-atoms were anti-ferromagnetically coupled to the Mn-atoms (Mn IIb) in the face centers with coordinates (0, 0.5, 0.5) and (0.5, 0, 0.5). Since the total atomic magnetic moments at all three Mn-sites do not cancel each other out (Mn I = 3.47 μ$_B$, Mn IIa = 0.75 μ$_B$ and Mn IIb = -2.36 μ$_B$,), the ground state is a ferrimagnet.

Table I lists the five samples (S1-S5) that were examined in this study. We begin by investigating different substrate annealing temperatures to understand the impact of the MgO surface on epitaxial growth of Mn$_4$N. The S1-S3 were deposited at 400 °C, and S4 and S5 were deposited at a higher temperature of 450 °C. Film thickness was determined by XRR measurement. In the case of S2-S5, the MgO substrates were pre-baked before loading into the sample deposition chamber.

TABLE I. List of Mn$_4$N thin films on MgO substrates

| Sample No. | MgO annealing Temperature | MgO annealing Time | Film thickness | Deposition temperature |
|---|---|---|---|---|
| S1 | No annealing (RT) | -- | 16.6 ± 0.6 nm | 400 °C |
| S2 | 1000 °C | 4 hours | 16.8 ± 0.5nm | 400 °C |
| S3 | 1100 °C | 4 hours | 16.8 ± 0.5 nm | 400 °C |
| S4 | 1100 °C | 4 hours | 16.8 ± 0.5 nm | 450 °C |
| S5 | 1100 °C | 4 hours | 11.5 ± 0.6 nm | 450 °C |

The film thicknesses measured by the XRR technique indicated that the thicknesses of all five thin film samples were less than 20 nm (see Table I). Analysis of XRR results is shown in Supplementary Figure 1. S1-S4 had a thickness of 16.8 ± 0.5nm, whereas S5 had a thickness of 11.5 ± 0.6 nm. Figure 2(a) shows the normalized M-H curves with out-of-plane applied magnetic field for 16.8 nm Mn$_4$N film on different heat-treated MgO substrates (S1-S3). The M-H curves show a strong dependence on the annealing temperature. All out-of-plane hysteresis loops were open, indicating PMA. The substrate annealing temperature of S3 (1100 °C) is higher than that of S2 (1000 °C), whereas S1 is unannealed. The squareness of the M(H) hysteresis loop improves from S1 to S2 to S3. Figure 2(b) shows the dependence of the substrate annealing temperature on $M_r/M_s$. The $M_r/M_s$ ratio increased from 0.47 to 0.72 as the annealing temperature increased from RT to 1100 °C. The increment of $M_r/M_s$ is attributed to the improved epitaxial growth of the thin film. High-temperature annealing not only decomposes the Mg(OH)$_2$ and MgCO$_3$ contaminants, it also reconstructs the surface of the MgO substrate [39,40]. During this process, the substrate forms an atomically smooth surface [39,40], which is beneficial for epitaxial growth [41]. The difference between S1 and S3 is mainly attributed to the different substrate annealing temperatures. Thin films annealed at 1100 °C produced a smoother surface than 1000 °C. The surface morphology as characterized by the atomic force microscopy revealed an average root-mean-square (rms) roughness of 0.592 nm, 0.308 nm, 0.278 nm for the as-received, 1000 °C annealed and 1100 °C annealed thin films, respectively (See Supplementary Figure 2).

To explore further improvement, we increased the deposition temperature to 450 °C, while retaining the substrate annealing temperature and time at 1100 °C and 4 hours. It has been shown that the squareness of the $Mn_4N$ film can be improved by increasing deposition temperature [29]. We found that the $M_r/M_s$ ratio of these samples, S4 and S5, increased further to near 1, as shown in Figure 2(c). The difference between S4 and S5 is in the film thickness: the S4 and S5 thin films were $16.8 \pm 0.5$ nm nm and $11.5 \pm 0.6$ nm thick, respectively (see Table I). S5, the thinnest film, kept a high $M_r/M_s$ ratio ~0.8, which is an encouraging outcome.

Figure 2(d) shows the out-of-plane and in-plane M-H curves for the S4 film. The $M_s$ and $K_u$ for the S4 film is $43 \pm 1$ emu/cc and 0.70 Merg/cc, respectively. All five samples had similar $M_s$ (40-60 emu/cc). Although the $M_s$ of these films are smaller than the reported $M_s$ (~100 emu/cc) in thicker films (30-100 nm) [22-25,28], they are comparable with that of the MBE-grown sub-20 nm $Mn_4N$ film on MgO, 50-80 emu/cc [26,27]. The lower $M_s$ of sub-20 nm $Mn_4N$ films may be due to surface oxidation [25].

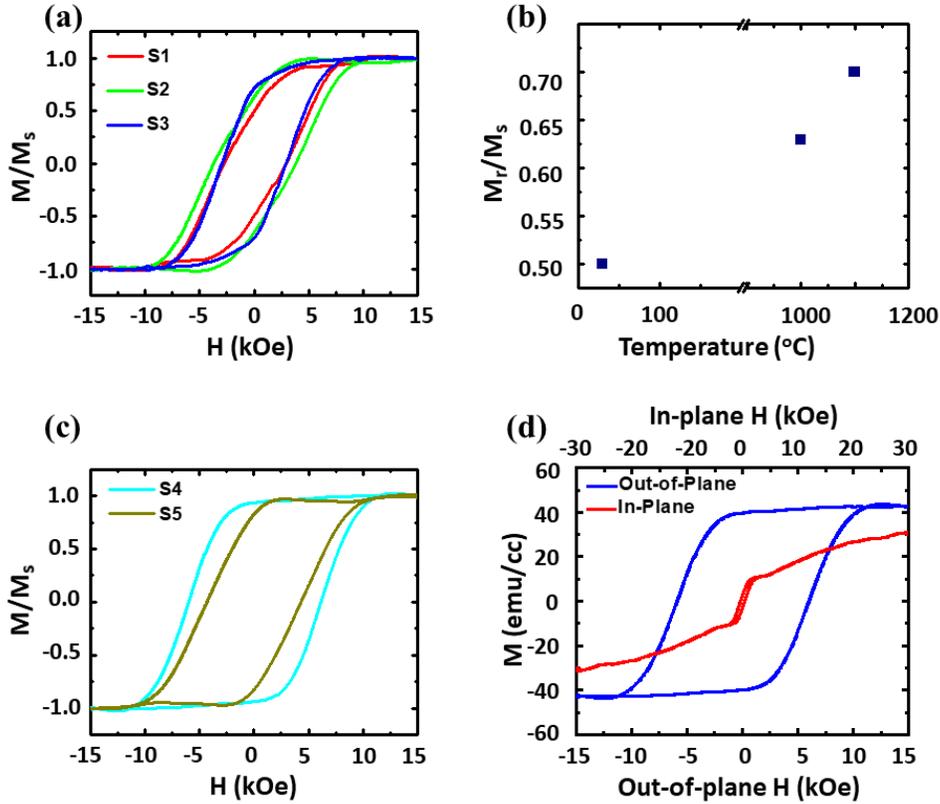

Figure 2 (a) Normalized out-of-plane M(H) loops of 16.8 nm $Mn_4N$ films deposited at 400 °C on MgO substrates annealed at different temperatures; S1 (red) was unannealed, S2 (green) was annealed at 1000 °C, and S3 (blue) was annealed at 1100 °C, (b) $M_r/M_s$ ratio of 16.8 nm $Mn_4N$ film (S1-S3) deposited at 400 °C on MgO substrates, which have been annealed at different temperatures, (c) Normalized out-of-plane M(H) loops of 16.8 nm (S4, cyan) and 11.5 nm (S5, brown) thick $Mn_4N$ films deposited at 450 °C. on MgO substrates, which have been annealed at 1100 °C , (d) out-of-plane and in-plane M(H) loops of the S4 $Mn_4N$ film.

Figure 3 shows the out-of-plane 2θ-θ XRD profile (a) and φ scan (b) of S3. Besides the MgO substrate peaks, only $Mn_4N$ (002) peak is observed in the 2θ-θ XRD profile, which indicates the $Mn_4N$ (00l) orientation is parallel to the MgO (00l). In the XRD 360° φ scan, both $Mn_4N$ and MgO shows four peaks with 90° interval, which corresponding to (202), (022), (-202), and (0-22). The overlap of $Mn_4N$ peaks and MgO peaks in φ-scan confirm their epitaxial relationship to be MgO (001)[100]// $Mn_4N$ (001)[100]. The out-of-plane lattice constant $c$ and in-plane lattice constant $a$ deduced from the (002) and (202) diffraction peaks are 0.386 nm and 0.391 nm, respectively. Therefore, the c/a ratio was calculated to be 0.987, which is close to the value reported earlier [22-28].

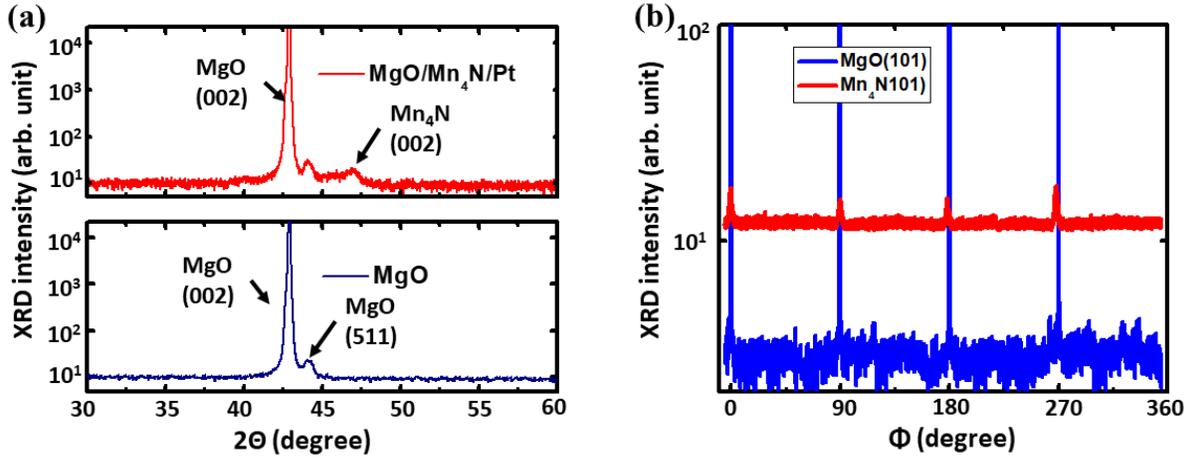

Figure 3 . (a) 2θ- θ profile of S3 and MgO substrate annealed at 1100 °C. (b) φ scan of Mn$_4$N (101) and MgO (101) peaks.

Based on the experimental lattice constant, DFT predicts an M$_s$ value of 153 emu/cc, which is larger than our experiment results. However, DFT is a zero-temperature electronic structure calculation. Li et al have shown that $M_s$ of the bulk Mn$_4$N would decrease as a function of increasing measurement temperature [40]. DFT also does not consider the possible effects of the mixing layer at the interface as well as defects. Future investigation will address the temperature dependence of saturation magnetization and magnetic anisotropy energy.

We have grown sub-20 nm ferrimagnetic Mn$_4$N epitaxial thin films with high M$_r$/M$_s$ ratios at 400-450 °C substrate temperatures on MgO substrates by reactive sputtering. The quality of the epitaxial growth was optimized by ex-situ pre-annealing of MgO substrates at temperatures as high as 1100 °C. The annealing was found to reduce the surface roughness of the MgO substrates at the atomic level, which was found to improve the quality of the out-of-plane magnetization hysteresis loops. The present results established Mn$_4$N thin film as a thermally stable ferrimagnet for further investigation as promising skyrmion-based spintronic material.

**Supplementary material**

See supplementary material for the XRR results of the film and the AFM results of the MgO substrates surface.

**Acknowledgment**

This work was supported by the DARPA Topological Excitations in Electronics (TEE) program (grant D18AP00009). The content of the information does not necessarily reflect the position or the policy of the Government, and no official endorsement should be inferred. Approved for public release; distribution is unlimited.

**Data Availability Statements**

The data that support the findings of this study are available from the corresponding author upon reasonable request.

# Supplementary Material for Rare-earth free ferrimagnetic $Mn_4N$ sub-20 nm thin films as potential high- temperature spintronic material

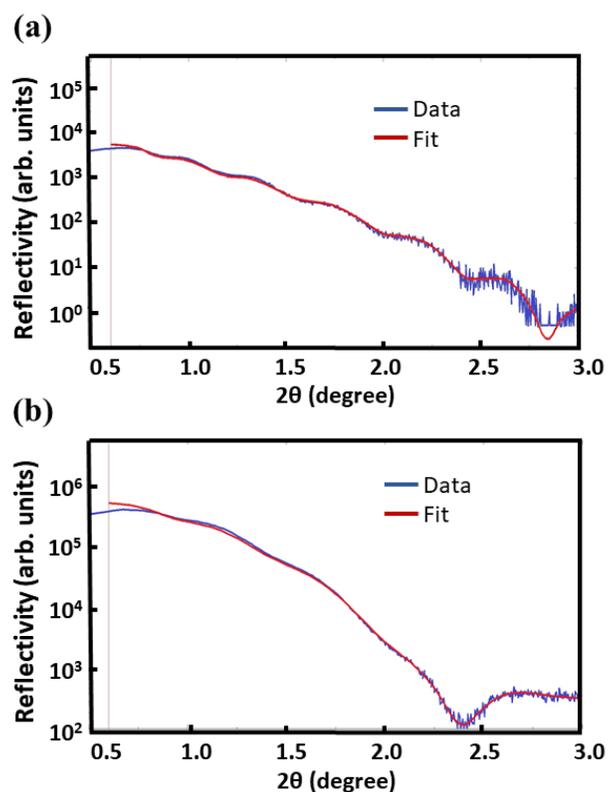

Supplementary Figure 1. XRR curve of the S3(a) and S5(b). the blue line is the raw data, and the red line is the fitting line.

Notes on Supplementary Figure 1.

X-ray reflectivity (XRR) was performed on the MgO/Mn$_4$N/Pt samples to confirm that the Mn$_4$N layer is thinner than 20 nm. The blue line is the raw data collected with a Cu-Kα source (λ = 1.5406 Å) and the red line is a fitting line produced by the software Rigaku-GXRR[31]. Supplementary Figure 1(a) shows the XRR curve and fitting line of S3. The fit gave a Mn$_4$N film thickness of 16.8 ± 0.5nm (nominal thickness was 15 nm) and a Pt capping layer thickness of 3.4 ± 0.6 nm nm. Supplementary Figure 1(b) shows the XRR curve and fitting line of S5. The fit gave a Mn$_4$N film thickness of 11.5 ± 0.6 nm (nominal thickness was 10 nm) and a Pt capping layer thickness of 4.2 ± 0.5 nm . The discrepancy between the XRR fitting model and the measured data at small angle (< 0.8) is because these angles are smaller than the critical angle (approx. 0.8 degree) for total reflection.

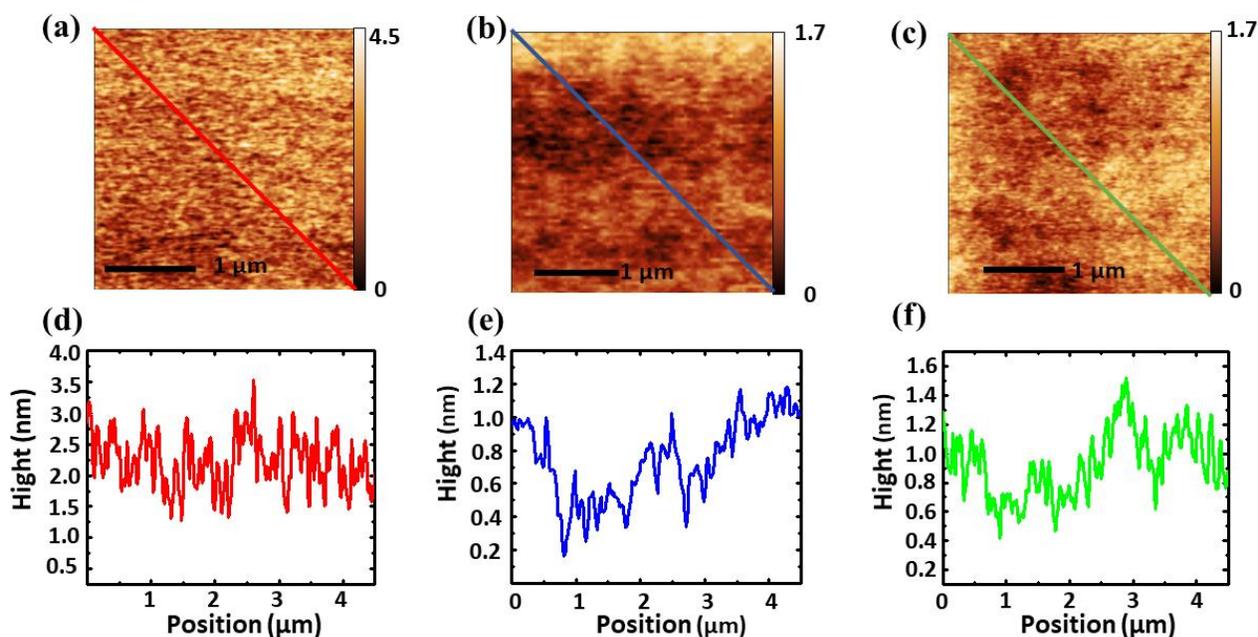

Supplementary Figure 2. AFM surface morphology of pure MgO (100) substrate surface (a) as received, and after annealing for 4 hours at (b) 1000 °C, (c) 1100 °C. The lower panels (d-f) show the corresponding line scans of the AFM topographies.

Notes on Supplementary Figure 2.

AFM was performed on the MgO substrates, which have been annealed at different temperatures, to check the surface morphology of the MgO (001) substrates. Supplementary Figure 2 (a), (b), (c) shows the AFM images of as-received MgO substrate, annealed at 1000 °C, and annealed at 1100 °C, respectively. Supplementary Figure 2 (d), (e), (f) shows the corresponding line scans of the AFM topographies. It is clear that as-received MgO has the roughest surface. However, from the images, it is hard to tell the difference between the 1000 °C annealed MgO substrate surface (Supplementary Figure 2(b), (e)) and the 1100 °C annealed MgO substrate surface (Supplementary Figure 2(c), (f)). Gwyddion [43] was employed to analyze the roughness, which gave the 1000 °C annealed MgO substrate surface rms roughness of 0.308 nm and the 1100 °C annealed MgO substrate surface rms roughness of 0.278 nm. For comparison, the surface rms roughness of the as-received substrate was 0.592 nm.